\documentclass[12pt]{article}
\usepackage[utf8]{inputenc}
\usepackage[usenames,dvipsnames]{color}
\usepackage{amsmath}
\usepackage{amssymb}
\usepackage{mathptmx}
\usepackage{authblk}

\usepackage[a4paper, margin=1in]{geometry}

\usepackage[breaklinks,colorlinks,citecolor=RoyalBlue,linkcolor=magenta]{hyperref}

\usepackage[super]{natbib}
\usepackage{graphicx}

\newcommand{\arcsec}{$^{\prime\prime}$}
\newcommand{\arcmin}{$^{\prime}$}
\newcommand{\arcdeg}{$^{\circ}$}
\newcommand{\micron}{$\mu$m}
\newcommand\farcs{\mbox{$.\!^{\prime\prime}$}}

\def\jnl@style{\it}
\def\aaref@jnl#1{{\jnl@style#1}}

\def\aj{\aaref@jnl{AJ}}                   % Astronomical Journal
\def\araa{\aaref@jnl{ARA\&A}}             % Annual Review of Astron and Astrophys
\def\apj{\aaref@jnl{ApJ}}                 % Astrophysical Journal
\def\apjl{\aaref@jnl{ApJ}}                % Astrophysical Journal, Letters
\def\apjs{\aaref@jnl{ApJS}}               % Astrophysical Journal, Supplement
\def\aap{\aaref@jnl{A\&A}}                % Astronomy and Astrophysics
\def\aapr{\aaref@jnl{A\&A~Rev.}}          % Astronomy and Astrophysics Reviews
\def\mnras{\aaref@jnl{MNRAS}}             % Monthly Notices of the RAS
\def\pasp{\aaref@jnl{PASP}}               % Publications of the ASP
\def\nat{\aaref@jnl{Nature}}              % Nature
\def\pasa{\aaref@jnl{PASA}}               % Publications of the Astron. Soc. of Australia
\def\ssr{\aaref@jnl{Space~Sci.~Rev.}}     % Space Science Reviews

\renewcommand{\figurename}{Fig.}

\newcommand{\mthc}{\mbox{CH$_3$CN}}
\newcommand{\mthtc}{\mbox{CH$_3$$^{13}$CN}}
\newcommand{\tmthc}{\mbox{$^{13}$CH$_3$CN}}
\newcommand{\mf}{\mbox{CH$_3$OCHO}}
\newcommand{\meth}{\mbox{CH$_3$OH}}
\newcommand{\fmh}{\mbox{H$_2$CO}}
\newcommand{\water}{\mbox{H$_2$O}}

\newcommand{\kms}{\mbox{km\,s$^{-1}$}}
\newcommand{\cc}{\mbox{cm$^{-3}$}}
\newcommand{\lsun}{\mbox{$L_\odot$}}
\newcommand{\msun}{\mbox{$M_\odot$}}

\newcommand{\hii}{\mbox{H\,{\sc ii}}}

\title{\textbf{A massive Keplerian protostellar disk with flyby-induced spirals in the Central Molecular Zone}}

\author[1,2*]{Xing Lu}
\affil[1]{\small Shanghai Astronomical Observatory, Chinese Academy of Sciences, 80 Nandan Road, Shanghai 200030, People’s Republic of China; \url{xinglu@shao.ac.cn}}
\affil[2]{\small National Astronomical Observatory of Japan, 2-21-1 Osawa, Mitaka, Tokyo 181-8588, Japan}

\author[3*]{Guang-Xing Li}
\affil[3]{South-Western Institute for Astronomy Research, Yunnan University, Kunming, 650500 Yunnan, People's Republic of China; \url{gxli@ynu.edu.cn}}

\author[4]{Qizhou Zhang}
\affil[4]{Center for Astrophysics $|$ Harvard \& Smithsonian, 60 Garden Street, Cambridge, MA 02138, USA}

\author[5]{Yuxin Lin}
\affil[5]{Max-Planck-Institut f{\"u}r Extraterrestrische Physik, Giessenbachstr.\ 1, D-85748 Garching bei M{\"u}nchen, Germany}

\date{}
 
\begin{document}

\maketitle

{\color{blue}
\noindent Accretion disks are an essential component in the paradigm of the formation of low-mass stars. Recent observations further identify disks surrounding low-mass pre-main-sequence stars perturbed by flybys. Whether disks around more massive stars evolve in a similar manner becomes an urgent question. We report the discovery of a Keplerian disk of a few solar masses surrounding a 32 solar-mass protostar in the Sagittarius C cloud around the Galactic Center. The disk is gravitationally stable with two embedded spirals. A combined analysis of analytical solutions and numerical simulations demonstrates that the most likely scenario to form the spirals is through external perturbations induced by a close flyby, and one such perturber with the expected parameters is identified. The massive, early O-type star embedded in this disk forms in a similar manner with respect to low-mass stars, in the sense of not only disk-mediated accretion, but also flyby-impacted disk evolution.
}

The Central Molecular Zone (CMZ) is a large molecular gas reservoir around the Galactic Center with unexpectedly inefficient star formation\citep{longmore2013,barnes2017,kauffmann2017a,lu2019b}. The existence of protostellar accretion disks in the CMZ has been indicated by the detection of presumably disk-driven protostellar outflows\citep{lu2021,walker2021}. However, direct imaging of disks has been challenging, since high angular resolutions are needed to resolve disks at the distance of the CMZ (8.1~kpc\citep{reid2019}).

\begin{figure}[!h]
\centering
\includegraphics[width=1\textwidth]{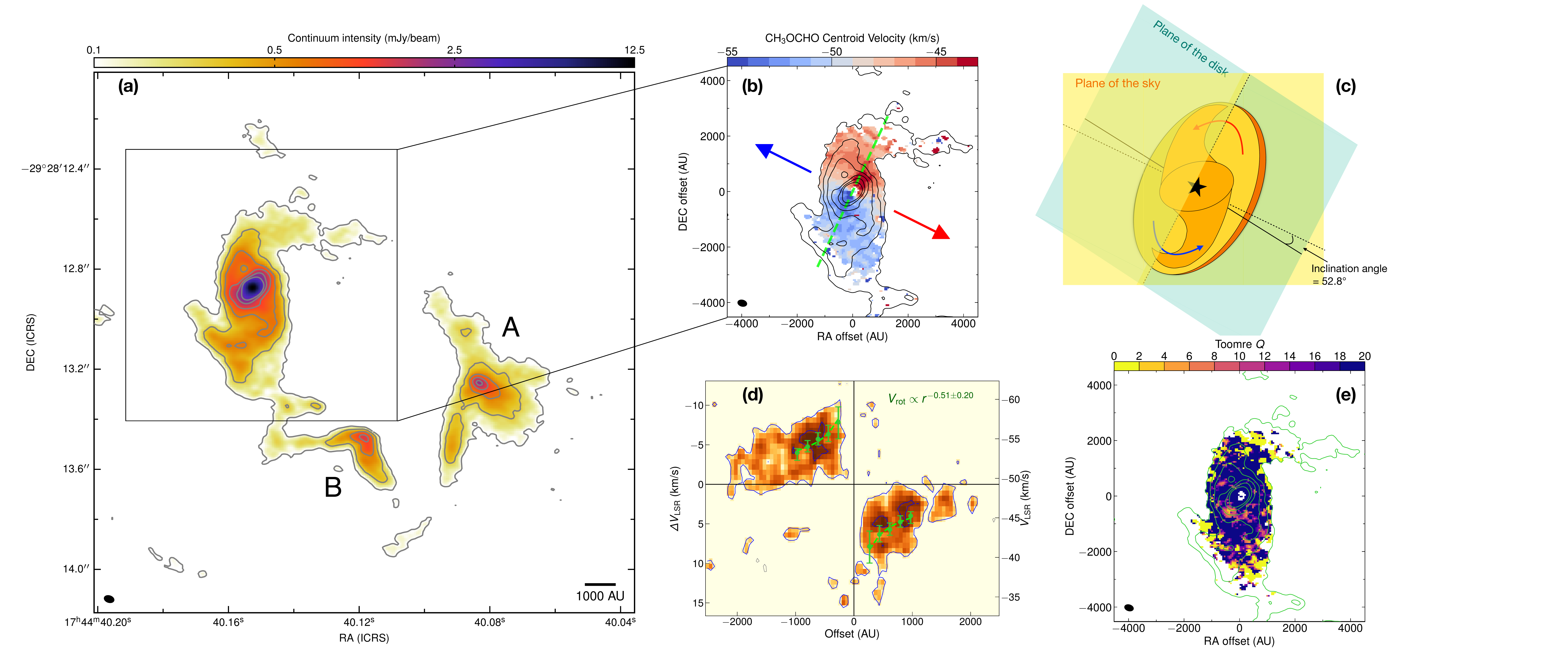}
\caption{
\textbf{Disk properties and configurations based on ALMA Band 6 (1.3~mm) observations.} \textbf{a.} Overview of the continuum emission. The contour levels are [5,15,35,65,105,155] $\times$ $\sigma$, where $\sigma=18~\mu$Jy per beam. \textbf{b.} Centroid velocities of the \mf{} molecular line. The contours are the same as in panel a. The blue and red arrows show the directions of the outflow seen in SiO emission in lower resolution data\citep{lu2021}. The green dashed line shows the best-fit kinematic major axis of the disk, with a position angle of 335\arcdeg{} (north through east). \textbf{c.} A schematic illustration of the three-dimensional configuration of the disk. The disk is inclined to 52.8\arcdeg{} so that it is slightly closer to being edge-on than face-on. \textbf{d.} The position-velocity diagram along the major kinematic axis of the disk as defined in panel b. The green dots with errorbars show the best-fit rotation velocities at each radius assuming 3D concentric rings. The green dashed curve shows the power-law fit to the dots, leading to the relation presented in the plot. \textbf{e.} The Toomre $Q$ parameters in the disk that take micro-turbulence into account. The image scale is truncated at a value of 20 to facilitate illustration. The map center in \textbf{b} \& \textbf{e}, where the continuum emission peak locates, is at ($\alpha$(ICRS), $\delta$(ICRS)) = 17h44m40.153s, $-$29\arcdeg{}28\arcmin{}12\farcs{876}.
}
\label{fig:disk}
\end{figure}

We report the direct imaging of a Keplerian protostellar disk in the CMZ by using long-baseline Atacama Large Millimeter/submillimeter Array (ALMA) observations. The target locates in the Sagittarius C cloud and shows a powerful outflow\citep{lu2021}. The disk is detected in the Band 6 (1.3~mm) dust continuum and molecular line emission (Fig.~\ref{fig:disk}), and extends to a radius of about 2000~AU. The disk mass amounts to 4.7~\msun{} based on the thermal dust emission, with an uncertainty of a factor of 2 (\hyperref[sec:methods]{Methods}). In the following, we refer to it as the Sgr~C disk, or simply the disk.

The line-of-sight gas velocities in the disk present a gradient in the northwest-southeast direction (Fig.~\ref{fig:disk}b), which is perpendicular to the major axis of the outflow (Extended Data Fig.~\ref{fig:largeenv}). Based on the gas velocities in the disk and in the outflow, we are able to constrain the three-dimensional configuration of the system (Fig.~\ref{fig:disk}c). The disk is best fit with an inclination angle of 52.8\arcdeg{} (0\arcdeg{} for face-on; \hyperref[sec:methods]{Methods}). The rotation velocities in the disk are well correlated with the disk radii, with a power-law index of $-0.51\pm0.20$ within a radius of 1000~AU (Fig.~\ref{fig:disk}d; \hyperref[sec:methods]{Methods}). This is consistent with the Keplerian rotation, i.e., when the rotation is controlled by the gravity of a central massive object such that the rotation velocities are proportional to the inverse square root of disk radii.

The protostellar mass, as determined by fitting a Keplerian model to the rotation velocities in the disk (\hyperref[sec:methods]{Methods}), is $31.7\pm4.7$~\msun{}. This is consistent with the expected mass from the luminosity of the central object, a few times 10$^5$~\lsun{}, estimated from the temperature profile (\hyperref[sec:methods]{Methods} and Extended Data Fig.~\ref{fig:tgas}). The protostar is among the most massive ones with disks reported in the literature\citep{beltran2016,zhao2020}.

Inside the disk, we find two spirals in both thermal dust and spectral line emission, which are trailing based on the velocities (Fig.~\ref{fig:disk}c).

\vspace{0.5em}
\noindent{\large \textbf{Discussion}}

\begin{figure}[!h]
\centering
\includegraphics[width=0.5\textwidth]{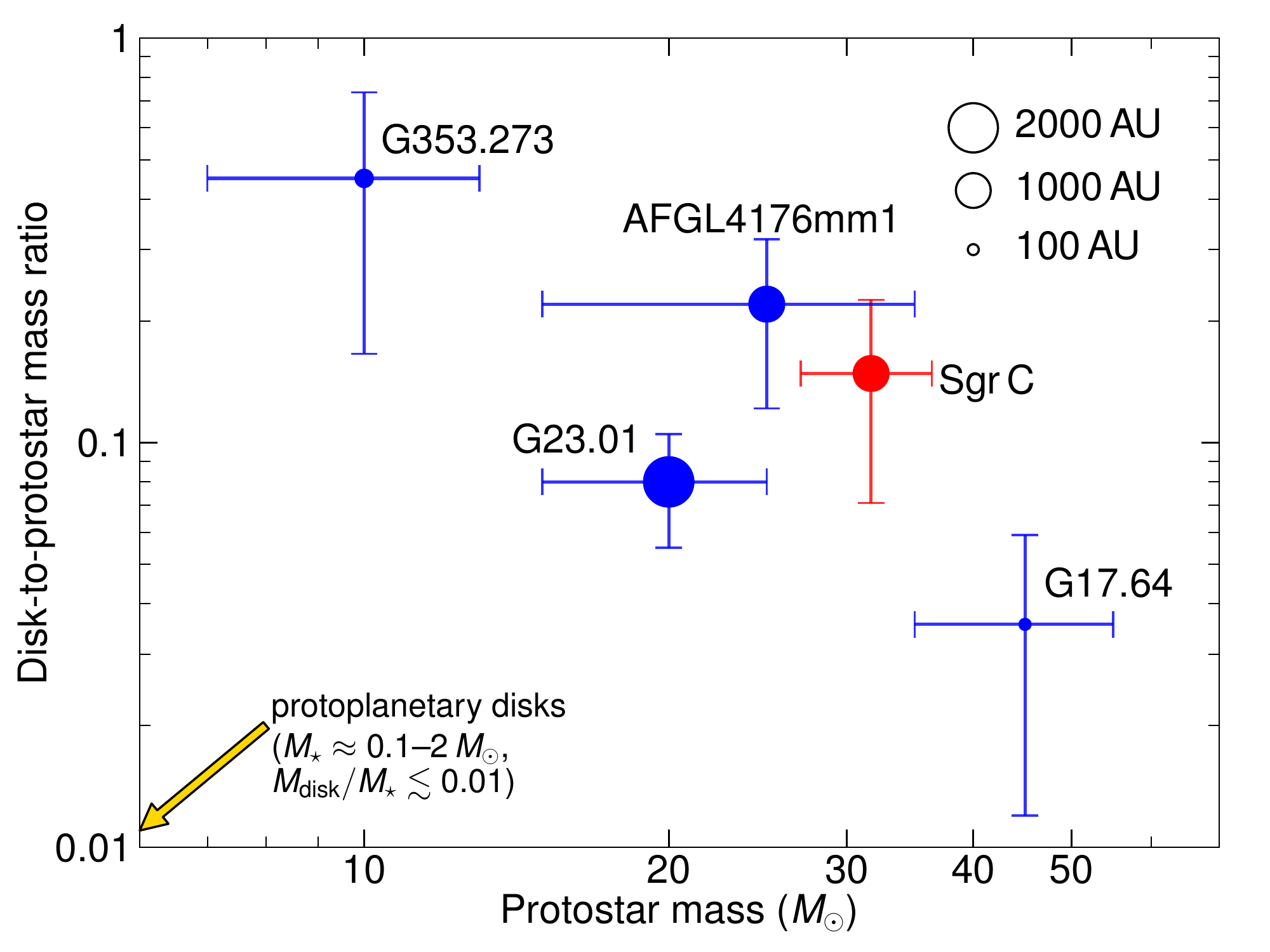}
\caption{
\textbf{Disk-to-protostar mass ratios versus protostar masses for spatially resolved massive protostellar disks with substructures.} The Sgr~C disk is highlighted with a red dot. The G17.64 disk can be gravitationally stable with Toomre $Q$ parameters $>$2, but it is highly dependent on temperature assumptions in the disk. The other three disks are suggested to be gravitationally unstable. Errorbars are compiled from the errors reported in respective references. Sizes of the dots correspond to the radii of the disks. In the bottom left corner, the population of protoplanetary disks around low-mass stars, which lie beyond the plotted range, are indicated with an arrow\citep{andrews2013,garufi2018}. References of the disks: AFGL\,4176\,mm1\citep{johnston2020}; G17.64\citep{maud2019}; G23.01\citep{sanna2019}; G353.273\citep{motogi2019}.
}
\label{fig:disksample}
\end{figure}

\noindent{\textbf{A Keplerian disk around an early O-type protostar}}

Recent high-resolution ALMA observations have revealed the existence of accretion disks around massive ($>8$~\msun{}) protostars\citep{johnston2015,sanna2019,maud2019,motogi2019,zapata2019,johnston2020,sanna2021}, suggesting that massive stars form through disk-mediated accretion in a similar manner as low-mass stars\citep{shu1987}. However, whether disks exist around even more massive, early O-type protostars ($\gtrsim30$~\msun{}) is still debatable, and several candidates have been suggested to be gravitationally collapsing envelopes rather than rotationally supported disks\citep{cesaroni2017}.

Our finding lends support to the existence of disks around early O-type protostars. The ratio between the dynamical mass of the rotating structure and its gas mass is $\gtrsim$10, suggesting that the structure is centrifugally supported\citep{beltran2016}. The Sgr~C disk is thus a true Keplerian disk around a 32~\msun{} protostar.

\noindent{\textbf{A gravitationally stable disk with spirals}}

A limited sample of spatially resolved observations have suggested that massive disks may contain various substructures such as spirals\citep{johnston2020,sanna2021}, rings\citep{maud2019}, and other asymmetric features\citep{motogi2019,zapata2019}. The substructures have been interpreted as fragments of the disks induced by gravitational instability\citep{motogi2019,johnston2020}, or as accretion streamers\citep{sanna2021}.

Gravitational instability is unlikely to be responsible for the spirals in the Sgr~C disk. The stability of disks can be characterized by the Toomre $Q$ parameter \citep{toomre1964}, where disks become unstable if $Q < 1\text{--}2$ \citep{durisen2007}. As shown in Fig.~\ref{fig:disk}e, the whole Sgr~C disk has sufficiently large $Q$ values ($Q \gg 2$), thus is a stable rotating structure insusceptible to gravitational collapse. Taking into account potential uncertainties in the $Q$ values does not change the conclusion (\hyperref[sec:methods]{Methods} and Extended Data Fig.~\ref{fig:toomreQ_more}). The accretion streamer scenario is inconsistent with the observations either. The spirals show a symmetric grand-design pattern in the dust emission and ordered rotation kinematics, which are not expected for gas streamers that are supposed to be irregularly asymmetric in morphology and kinematics\citep{sanna2021}.

We compare the Sgr~C disk to a sample of spatially resolved massive protostellar disks with substructures in Fig.~\ref{fig:disksample}. Only four other sources are found in the literature, due to the rarity of massive stars as well as the difficulty of obtaining high-resolution data to resolve the disks. Among the five disks, the Sgr~C disk stands out as the only one confirmed to be gravitationally stable. Compared to protoplanetary disks around low-mass stars, disks around mass protostars have high disk-to-protostar mass ratios, indicating early evolutionary stages. The Sgr~C disk thus occupies the high protostar mass and high disk-to-protostar mass ratio end of the parameter space.

\begin{figure}[!t]
\centering
\includegraphics[width=0.75\textwidth]{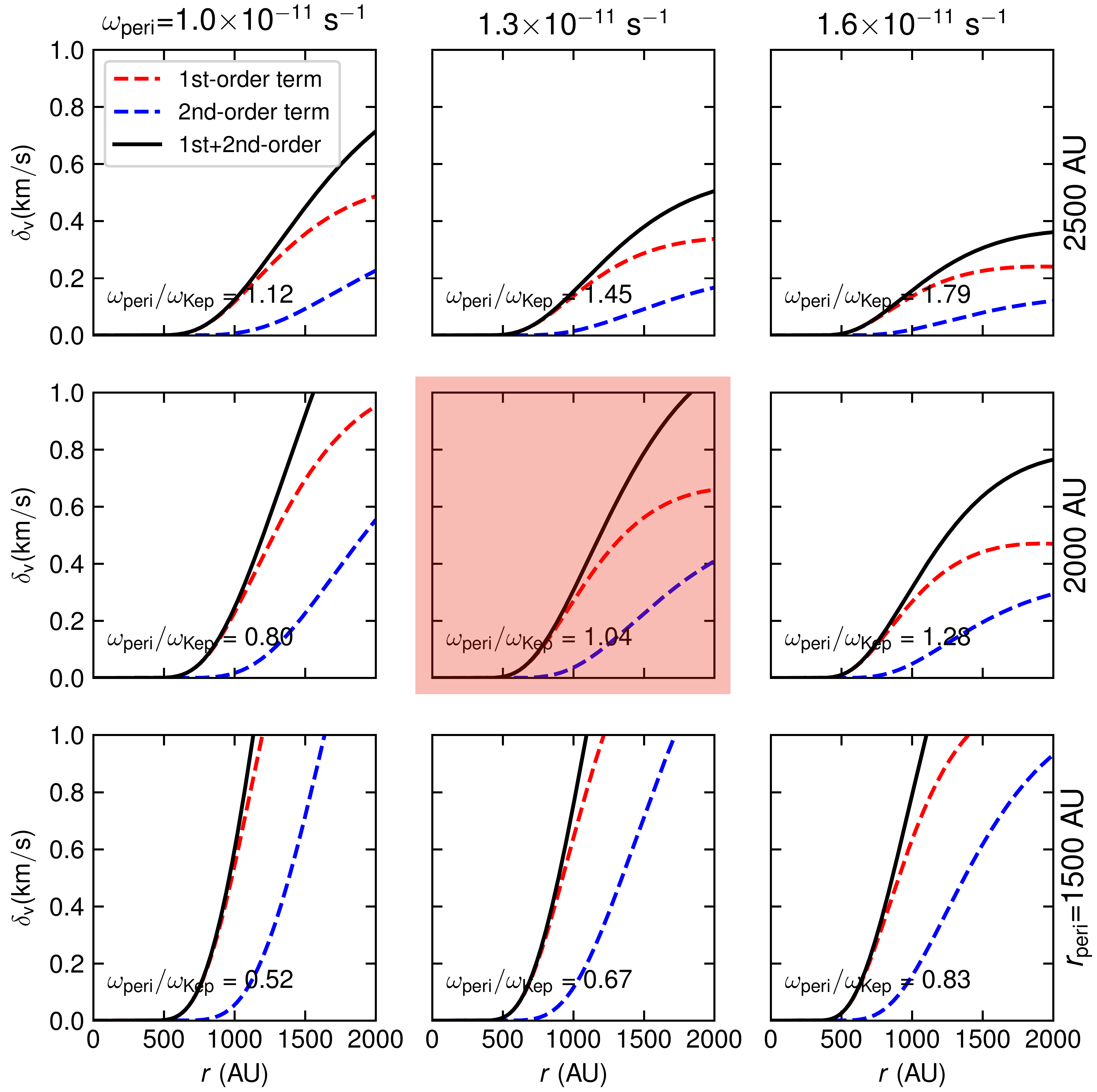}
\caption{
\textbf{Parameter space of perturbation-induced structure formation.} Panels show the velocity response of the disk from flyby events of different angular velocities at the periastron ($\omega_\text{peri}$) and different periastron radii ($r_\text{peri}$). The red and blue dashed lines represent contributions from the first and second-order terms, respectively\citep{donghia2010}, while the sums of the two terms are indicated by the black solid lines. Ratios between $\omega_\text{peri}$ and the Keplerian angular velocity at the periastron, $\omega_\text{Kep}$, are labelled. The highlighted panel is where the velocity response in the disk between radii of 1000-2000~AU is $\sim$1~\kms{}. A more extensive parameter space is shown in Extended Data Fig.~\ref{fig:pd_more}.
}
\label{fig:pd}
\end{figure}

\noindent{\textbf{A close flyby triggering the formation of the spirals}}

Since the disk is stable against self-gravity, an alternative explanation is that the spirals emerge due to perturbation\citep{clarke1993}. The source of the perturbation can be internal or external: in the former case, protostellar companions or planets embedded in a disk could disturb the gas and dust and result in spirals, while in the latter case, stellar companions or gas condensations could pass by and perturb an initially stable disk through resonance, creating substructures. These mechanisms have been well understood in the context of protoplanetary disk evolution\citep{pfalzner2003,bate2003,xiang2016,vincke2016,winter2018,cuello2019}, and several candidate cases have been observed \citep{cabrit2006,dai2015,kurtovic2018,rodriguez2018,akiyama2019,perez2020,menard2020,zapata2020,dong2022}.

We suggest that an external perturbation is the most likely origin of the spirals. The disk is not isolated. Next to the disk at a projected distance of about 8000 AU is a gas condensation traced by the continuum emission, labelled as A in Fig.~\ref{fig:disk}. The mass is 3.2~\msun{} with an uncertainty of a factor of 2 (\hyperref[sec:methods]{Methods}). The detection of a candidate bipolar outflow (Extended Data Fig.~\ref{fig:largeenv}) indicates its protostellar nature. There is a less massive condensation B with 1.3~\msun{}, which should have weaker effects on the disk, and thus we do not discuss it further.

We first determine that condensation A is capable of perturbing the disk through an analytical approach (\hyperref[sec:methods]{Methods}). The momentum injection from the perturber into the disk should be sufficiently strong to induce a considerable velocity difference---ideally comparable to the sound speed in the disk ($\sim$1~\kms{}) to impact gas dynamics. As shown in Fig.~\ref{fig:pd}, to best reproduce the observed perturbed disk at radii of 1000--2000~AU, we require the angular velocity of the perturber similar to the Keplerian value (thus a resonance) and a periastron distance of $\approx2000$~AU.

In the coplanar case, the required angular velocity suggests a bound orbit. If we assume a parabolic orbit for the perturber whose angular velocity at the periastron should be $\sqrt{2}$ times the Keplerian value, then to reconcile with the result in Fig.~\ref{fig:pd}, a non-coplanar flyby with an inclination angle of 45\arcdeg{} is necessary so that the projected angular velocity onto the disk plane becomes similar to the Keplerian value.

\begin{figure}[!t]
\centering
\includegraphics[width=1\textwidth]{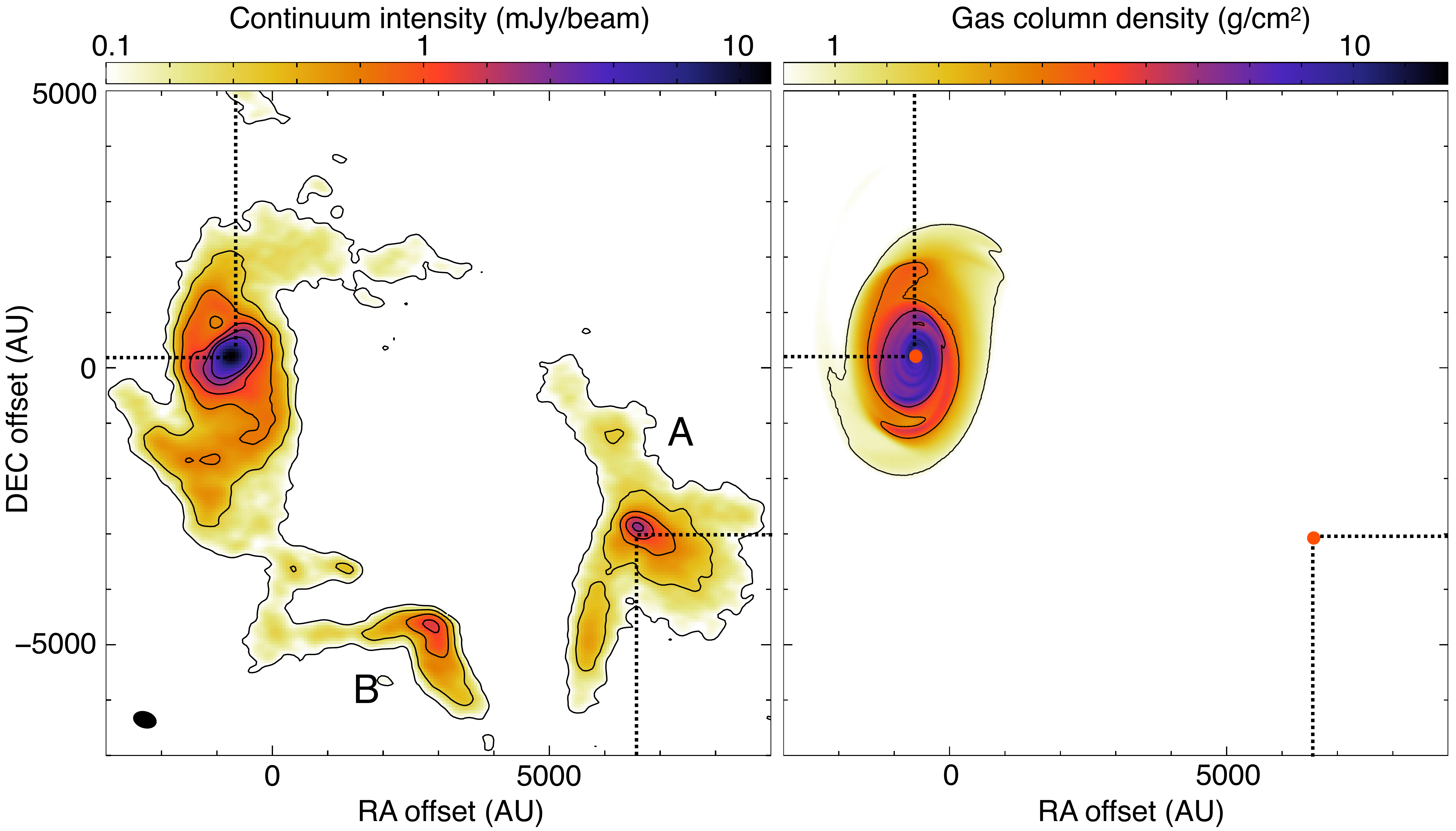}
\caption{
\textbf{Comparison between the observed environment of the disk and the simulated system after a flyby event.} \textit{Left}: Observed dust continuum emission. The contour levels are the same as in Fig.~\ref{fig:disk}. Condensations A and B are labelled. \textit{Right}: Gas column density map from the numerical simulation, taken at about 12,600 years after the flyby event. The system has been rotated to match the position angle and inclination angle of the observed disk. The contours are at [0.9, 2.6, 4.5]~g\,cm$^{-2}$. The two sink particles, representing the massive protostar embedded in the disk and condensation A, respectively, are marked by red dots. In both panels, the horizontal and vertical dashed lines indicate the positions of the sink particles.
}
\label{fig:flyby}
\end{figure}

As the flyby parameters are settled, we run a numerical simulation to verify our analytical solutions (\hyperref[sec:methods]{Methods}), and find that the simulated system reproduces all essential dynamical features of the observations. The right panel of Fig.~\ref{fig:flyby} shows the simulated gas column density map. The morphology and phase of the spirals in the disk, the tidal tail that points to condensation A, and the location of condensation A bear a clear resemblance to the observed image in the left panel.

We additionally find a few observational facts as \textit{a posteriori} evidence of the flyby scenario. The disk spatially coincides with an extended green object \citep{chambers2011,kendrew2013}, a class of infrared sources that are associated with active outflow and accretion activities. Multiple types of masers (\water{}, \meth{}, \fmh{}), indicating outflows and therefore active accretion events, have been detected toward this source\citep{lu2019a,lu2019b}. These observations indicate past episodes of active accretion that can be the consequences of flyby-induced perturbations \citep{cuello2019,borchert2022}.

\vspace{0.5em}
\noindent{\large \textbf{Perspectives}}

The discovery of the massive and perturbed Sgr C disk provides evidence that massive OB type stars may form in a similar way with low-mass stars. The existence of the Sgr C disk favors a scenario where early O-type stars accumulate mass through accretion disks analogous to low-mass stars\citep{beltran2016,zhao2020}. The gravitationally stable disk provides a unique sample to examine the mechanism to form spirals through flybys---a picture that has been acknowledged in the evolution of protoplanetary disks around low-mass stars\citep{pfalzner2003,cuello2019}. The flybys can trigger the formation of disk substructures, which might lead to the formation of stellar companions\citep{clarke1993}, a reminiscence of the cases toward low-mass multiple stellar systems formed via disk fragmentation\citep{tobin2016}. We expect the picture of disk-mediated accretion and flyby-impacted disk evolution toward massive protostars to be confirmed with future observations toward various regions, especially toward those embedded in forming massive clusters and with spirals in their disks\citep{johnston2020,sanna2021}.

\clearpage

\phantomsection
\label{sec:methods}
\noindent{\large \textbf{Methods}}

\noindent\textbf{Observations and data reduction}

The ALMA 12-m array observations toward the Sagittarius C region were carried out as part of a Cycle 6 project in five epochs on June 5, and July 14 and 15, 2019. The correlators were tuned to Band 6 (at the wavelength of 1.3 mm) to cover several molecular lines tracing protostellar disks, including \mthc{} and its isotopologues \mthtc{} and \tmthc{}. A large collection of complex organic molecules, including the \mf{} presented here, are also covered. Between 42 and 48 operational 12-m antennas were in the array during the five epochs of observations. The baseline lengths range from 83~m to 15238~m. The quasar J1924$-$2914 was used for flux and bandpass calibration, and J1744$-$3116 for phase calibration. The total on-source time toward the Sagittarius C region is 64.6~minutes. The pointing center is ($\alpha$(ICRS), $\delta$(ICRS)) = 17h44m40.37s, $-$29\arcdeg{}28\arcmin{}14\farcs{79}, which is offset from the continuum emission peak in Fig.~\ref{fig:disk} in order to simultaneously cover another target in the region.

The visibility data were calibrated using a pipeline in CASA version 5.4.0. Then we performed self-calibration and imaging using CASA version 5.6.1. We first made continuum-included spectral line dirty image cubes, with which line-free channels were visually identified. Next, we performed self-calibration using the continuum emission averaged from the line-free channels, for each of the five epochs. The self-calibration solutions were applied to the five continuum datasets respectively, but not to the spectral line data, as the latter are not dynamic range limited and therefore the improvement by applying self-calibration is minimal. The continuum data from the five epochs were combined and imaged using the TCLEAN task, with a Briggs weighting parameter of 0.5. The multiscale multifrequency synthesis algorithm was employed, with multiscale parameters of [0,5,15,50] times the pixel size of 0\farcs{007}. The final synthesized beam (angular resolution) is 0\farcs{043} $\times$ 0\farcs{030} with a position angle of 72.75\arcdeg{}. This beam size is equivalent to 350~AU $\times$ 240~AU, with the distance of 8.1~kpc to the Galactic Center. The image rms noise is 18~$\mu$Jy per beam, leading to a dynamic range of 700 given the peak intensity of 12.5 mJy per beam. The FWHM primary beam size (field of view) is 26\arcsec{}. The largest recoverable angular scale is 0\farcs{7}, as determined by the short baselines in the array.

The spectral lines were also imaged using the same parameters, with a channel width of 0.33~\kms{} for the \mthc{} and \mthtc{} lines, and 0.67~\kms{} for the other lines including \tmthc{} and \mf{}. The image rms noise is 1.2 or 0.8 mJy per beam, respectively, for the two channel widths. Finally, the continuum and spectral line images were corrected for the primary beam attenuation.

\vspace{0.2em}
\noindent\textbf{Distance and environment of the target}

The observed target is spatially projected inside the Sagittarius C molecular cloud. The location of the Sagittarius C cloud being within the CMZ is well established, given its physical associations with adjacent \hii{} regions and consistency with the overall gas kinematics in the CMZ \citep{kendrew2013,kruijssen2015,lu2019a}. Whether the disk is embedded in the Sagittarius~C cloud, however, needs discussion.

We have several reasons to believe that the disk is indeed in the Sagittarius C cloud therefore at a distance of about 8.1~kpc. First, its LSR velocity is at around $-$50~\kms{}, which is consistent with the velocity of the Sagittarius C cloud as well as the overall gas kinematics in the CMZ \citep{molinari2011,kruijssen2015,lu2019a}. Second, the velocity dispersion of molecular lines is typically $>$2~\kms{}, which is much broader than that seen in the Galactic disk but resembles the characteristic value in the CMZ.

The Sagittarius C cloud itself is an active star forming region, with a dozen of masers detected in previous observations \citep{lu2019a,lu2019b}. The targeted region is one of the confirmed high-mass star forming sites in the cloud, showing various masers, active outflow activities, and bright infrared emission \citep{kendrew2013,lu2019a,lu2019b,lu2021}. The 1.3~mm continuum flux within a radius of 0.07~parsec, obtained by the Submillimeter Array, is about 0.3~mJy\citep{lu2019a}. The continuum flux within the same area recovered by our ALMA observations is 0.14~mJy. Therefore, about half of the continuum emission from predominantly diffuse gas has been filtered out in our ALMA interferometer observations.

Previous radio continuum observations reveal an ultra-compact \hii{} (UC\hii{}) region to the south of the disk\citep{lu2019a,lu2019b}. The two independent VLA observations reach a consistent location for the UC\hii{} region, which are shown as contours in Extended Data Fig.~\ref{fig:largeenv}. The UC\hii{} region is unsolved or marginally resolved in the VLA observations, suggesting that its size may be smaller than the synthesized beam (down to 1\arcsec{}), while the projected angular distance between its peak and the disk is more than 2\arcsec{}. In addition, our recent ALMA 3-mm continuum observations resolve two components in this field, one coincident with the UC\hii{} region, and the other with the disk. Therefore, the UC\hii{} region is likely external to the disk system. It is probably much fainter at 1.3~mm or spatially resolved out in our ALMA longbaseline observation, and its impact on the disk is yet to be explored.

\vspace{0.2em}
\noindent\textbf{Estimate of gas temperatures}

Gas temperatures in the disk are estimated using two groups of molecular lines, the $K$-ladder of \mthc{} $J=12\text{--}11$ transition, which is around the rest frequency of 220.7~GHz, and that of \tmthc{} $J=13\text{--}12$ transition, around the rest frequency of 232.2~GHz. The \mthc{} lines are largely blended with transitions of another isotopologue, \mthtc{} $J=12\text{--}11$, as well as many other molecular lines in the same frequency range. We manage to simultaneously fit \mthc{}, \mthtc{}, and several commonly seen molecular lines (e.g., \meth{}, CH$_3$CH$_2$OH, HNCO) in the frequency range, assuming local thermodynamic equilibrium, with the XCLASS package\citep{moller2017}. Similarly, we include several commonly seen molecular lines in the fitting for \tmthc{}. The purpose of including the other molecular lines is to subtract them from the spectra and prevent them from interfering with the fitting of the \mthc{} and \tmthc{} lines. 

We assume that the molecular lines under consideration do not have very different velocities, and allow the centroid velocities of all the lines to vary within the range of $\pm$4~\kms{} with respect to the centroid velocity of the \mf{} line. Here the centroid velocity of \mf{} is used as the reference because this line is largely free of line blending as discussed below. The upper limit on all the FWHM linewidths is 6~\kms{}. Constraints on the centroid velocities and linewidths help converge the multi-line fitting in such frequency ranges of high spectral confusion. We also exclude the $K<4$ ladders in regions where these lower energy transitions become optically thick, i.e., where self absorption or saturated emission shows up.

The rotational temperatures for \mthc{} and \tmthc{} are derived in a pixel-by-pixel manner to obtain the temperature maps, as presented in Extended Data Fig.~\ref{fig:tgas}. It can be clearly seen that the temperatures based on \mthc{} are systematically higher than those from \tmthc{}. This is because the \mthc{} lines have a higher optical depth, therefore preferably trace the surface of the disk, which is likely externally heated by the protostar(s) at the center. The \tmthc{} lines, which are optically thinner, thus better trace the interior of the disk where the bulk of gas is not affected by external heating. As such, we use the temperature map based on \tmthc{} to estimate the mass and the Toomre $Q$ parameter of the disk in the following.

\vspace{0.2em}
\noindent\textbf{Properties of the central protostar: luminosity}

Once the spatial distribution of temperatures is derived, we are able to estimate the luminosity of the central heating source, presumably a massive protostar, using the Stefan-Boltzmann law:
\begin{equation}\label{equ:sblaw}
L=4\pi r^2 \sigma T^4,
\end{equation}
in which $T$ is the temperature at the radius $r$, and $\sigma$ is the Stefan-Boltzmann constant. As shown in the temperature map derived from \tmthc{} in Extended Data Fig.~\ref{fig:tgas}, at the center of the disk, within a radius of 200~AU (corresponding to an angular radius of 0\farcs{025}), the mean temperature is 800~K. The luminosity is then $7\times10^5$~\lsun{}. This is the expected luminosity of an O5-type zero-age main sequence star\citep{panagia1973}, whose mass would be about 40~\msun{}.

The standard deviation of the temperatures within the radius of 200~AU is about 500~K. We adopt a lower limit of 300 K for the temperature, and estimate a luminosity of $1.5\times10^4$~\lsun{}, corresponding to a B0-type zero-age main sequence star of 15~\msun{}. This is still qualified as a massive protostar.

We note that beyond the simple Stefan-Boltzmann law, previous studies have taken dust emissivity into account to derive a revised relation between the temperature profile and the luminosity\citep{scoville1976,garay1999}. For example, the following relation is from Ref.\citep{garay1999}:
\begin{equation}\label{equ:sblaw_garay}
L=\left[\left(\frac{T_D}{65~\text{K}}\right)\left(\frac{0.1}{f}\right)^{-1/(4+\beta)}\left(\frac{0.1~\text{pc}}{r}\right)^{-2/(4+\beta)}\right]^{4+\beta}~10^5~\lsun,
\end{equation}
in which $T_D$ is the dust temperature at the radius $r$, $f$ is the dust emissivity at the wavelength of 50~\micron{} and usually adopted to be 0.1, and $\beta$ is the spectral index of dust emissivity at far-infrared wavelengths that is usually in the range of 0 to 1\citep{garay1999}. If we assume that $\beta=0$, then using the temperature of 800~K at the radius of 200~AU, the luminosity of the central protostar is estimated to be $2.2\times10^5$~\lsun{}. If a larger $\beta$ is assumed, the derived luminosity will be higher.

Finally, all the above relations assume a sphere instead of a disk. If a disk is considered, an even higher luminosity will be estimated, because more emission would have escaped freely from the poles. Therefore, a massive protostar with a luminosity of $\gtrsim10^5$~\lsun{} should exist at the center of the disk.

\vspace{0.2em}
\noindent\textbf{Properties of the central protostar: dynamical mass}

As the rotation of the inner disk is Keplerian, we are able to estimate the mass of the protostar based on dynamical arguments.

The disk kinematics are fit with the 3DBarolo code \citep{diteodoro2015}, which fits a 3D position-position-velocity image cube with a series of concentric tilted rings, and estimates the inclination angle of a disk by assuming a circular geometry.

We use the \mf{} line to trace the kinematics in the disk, because it is detected with high signal-to-noise ratios and is relatively isolated from other molecular lines thus suffering less line confusion. As the signal-to-noise ratios are reasonably high ($>$10), we do not constrain the initial model parameters other than limiting the rotation velocity to be lower than 20~\kms{}, and directly feed the image cube into the code to obtain the best-fit model. 

The best-fit parameters of the disk are as follows: the inclination angle is 52.8\arcdeg{} (0\arcdeg{} being face-on), the position angle is about 335\arcdeg{} (north through east) with a scatter of a few degrees across different rings, and the systemic velocity is $-$49.2~\kms{}. The best-fit rotation velocities are plotted as dots in Fig.~\ref{fig:disk}d. The errorbars in the rotation velocities are estimated through a Monte-Carlo approach. A power-law fit to the data in the inner 1000~AU leads to the relation labelled in Fig.~\ref{fig:disk}d, $v_\text{rot}\propto r^{-0.51\pm0.20}$, which is consistent with Keplerian rotation.

We then estimate the mass of the central protostar embedded in the disk under the assumption of Keplerian rotation (i.e., exactly $v_\text{rot}\propto r^{-0.5}$). The mass is thus solely determined by the normalization of the power-law relation, which gives rise to a value of 31.7~\msun{}. The uncertainty in the mass, inherited from the errors in the rotation velocities, is 4.7~\msun{}.

\vspace{0.2em}
\noindent\textbf{Masses of the disk and the two condensations}

The gas mass in the disk is estimated using the continuum emission. As discussed above, the observed 1.3~mm continuum emission should be dominated by thermal dust emission except at the disk center. At the gas densities typically found in accretion disks ($\gtrsim10^7$~\cc{}), the gas and dust are well mixed and in thermal equilibrium. Therefore, we use the gas temperature to represent the dust temperature. The brightness temperatures of the dust emission in the disk are usually $\lesssim30$~K, much lower than the gas temperatures derived above. As such, the dust emission should be optically thin.

We estimate the gas mass in the disk following
\begin{equation}\label{equ:mdisk}
M=g\frac{S_\nu d^2}{B_\nu(T)\kappa_\nu},
\end{equation}
in which $g$ is the gas-to-dust mass ratio, $S_\nu$ is the dust emission flux, $d$ is the distance of 8.1~kpc, $B_\nu(T)$ is the Planck function at the dust temperature $T$, and $\kappa_\nu$ is the dust opacity. We adopted $g = 100$ and $\kappa_\nu = 1.99$~cm$^2$\,g$^{-1}$ (MRN model without ice mantles for gas density of $10^6$~\cc{}\citep{OH1994}), respectively.

The total continuum emission flux in the disk is measured to be 86~mJy. However, the central component is likely dominated by free-free emission around the massive protostar. We therefore first subtract a Gaussian whose peak equals the peak intensity and whose shape is the same as the synthesized the beam from the disk center, and then consider the residual flux of 76~mJy. The disk mass then amounts to 4.7~\msun{}.

The masses of the two condensations are estimated following the same approach, as the continuum is dominated by thermal dust emission as well. However, in the two condensations, the signal-to-noise ratios of the \mthc{} and isotopologues lines are not sufficient to allow temperatures to be determined. Instead, we use the upper energy level of the \mf{} transition detected in the condensations, 111.5~K, as an approximate of the temperature. Then the masses of condensations A and B were estimated to be 3.2 and 1.3~\msun{}, respectively. The actual kinetic temperature can be higher or lower than the upper energy level. If we assume an uncertainty of 50\% for the temperature, the uncertainty of the condensation masses would be about a factor of 2.

\vspace{0.2em}
\noindent\textbf{Estimate of the Toomre $Q$ parameter}

The Toomre $Q$ parameter is derived as
\begin{equation}\label{equ:toomreq}
Q=\frac{c_s\kappa}{\pi G\Sigma},
\end{equation}
where the sound speed $c_s$ and the surface density $\Sigma$ are defined as in previous sections, and $G$ is the gravitational constant. The epicyclic frequency, $\kappa$, is equivalent to the angular velocity for Keplerian rotation. Then the Toomre $Q$ profile is the disk can be calculated using these parameters, as displayed in Extended Data Fig.~\ref{fig:toomreQ_more}.

We also take the support of micro-turbulence against self-gravity into account, by substituting the sound speed in Equ.~\ref{equ:toomreq} with the total velocity dispersion, including both the sound speed and the micro-turbulent velocity dispersion, the latter estimated from the \mf{} line. The result has been shown in Fig.~\ref{fig:disk}e.

The uncertainty in the estimated Toomre $Q$ parameters is inherited from those of three factors: the sound speed (or velocity dispersion), the surface density, and the epicyclic frequency. In the case that includes micro-turbulence (Fig.~\ref{fig:disk}e), the uncertainty of the total velocity dispersion is dominated by that of the fitted velocity dispersion of \mf{}. The measured velocity dispersions are sufficiently broad, equivalent to at least 5 channels. The signal-to-noise ratios of the \mf{} line peak within the disk are 5--10. Therefore, the uncertainties in the linewidth should be small, to which we assign a value of 20\%.

The uncertainty associated with the surface density has been outlined in Ref.\citep{lu2019a}. The uncertainty is dominated by those of the dust opacity, the gas-to-dust ratio, and the temperature.  Following Ref.\citep{lu2019a}, we adopt uncertainties of 28\%, 50\%, and 50\%, respectively, for the three factors. Then the uncertainty in the surface density is 76\%.

The epicyclic frequency has been assumed to be equal to the angular velocity of Keplerian rotation. However, the disk rotation deviates from Keplerian especially in the outer part, as can be seen in Fig.~\ref{fig:disk}d. Therefore, we also directly calculate the epicyclic frequency $\kappa$:
\begin{equation}\label{equ:kappa}
\kappa^2=\frac{2\Omega}{r}\frac{\textrm{d}(r^2\Omega)}{\textrm{d}r},
\end{equation}
and obtain a Toomre $Q$ map. The result, as plotted in Extended Data Fig.~\ref{fig:toomreQ_more}, is consistent with that in Fig.~\ref{fig:disk}e: large $Q$ values of $\gg$2 are found throughout the disk.

Overall we estimate the uncertainty in the Toomre $Q$ parameters to be about 80\%. The lower limit of this range means that the Toomre $Q$ parameters could have been overestimated by a factor of 5. However, even if we decrease the values in Fig.~\ref{fig:disk}e by this factor, the Toomre $Q$ parameters throughout the disk are still $>$2. Therefore, the uncertainty in the Toomre $Q$ parameters does not affect our conclusion.

\vspace{0.2em}
\noindent\textbf{Analytical and numerical solutions of the flyby event}

We determine the criteria for a perturber to leave significant impacts on the disk dynamics. As discussed in the main text, the velocity change in the disk induced by flyby should be comparable to the sound speed, $\sim$1~\kms{}. If the change is too high, the disk would be strongly disturbed and thus truncated. If too low, the disk would not be significantly disturbed, and spirals would not emerge.

We then use the mass of condensation A as an input and compute the response of the disk with a set of different angular velocities ($\omega_\text{peri}$) and periastron radii ($r_\text{peri}$) using the formalism presented in Ref.\citep{donghia2010}. These are derived under the quasi-resonant condition. We search the parameter space extensively (Extended Data Fig.~\ref{fig:pd_more}) and find that the highlighted one in Fig.~\ref{fig:pd} remains to be the most suitable flyby parameter set.

Then we feed this parameter set into the numerical simulation. We use the smoothed particle hydrodynamics code \texttt{PHANTOM}\citep{price2018}, which has been used to model flyby events in several previous works\citep{cuello2019,menard2020,dong2022}. As explained in the main text, the parabolic flyby orbit has to be inclined by 45\arcdeg{} with respect to the disk plane. The flyby is set to be prograde, i.e., the projected velocity of the perturber on the disk plane being in the same direction as disk rotation at the periastron. We also consider that the initial size of the disk must be somewhat larger than what is observed, because close flybys truncate disks to smaller sizes\citep{breslau2014,bhandare2016}. Therefore, the disk outer radius is set to 3000~AU. The perturber thus penetrates the disk during the flyby. The inner radius is set to 50~AU, allowing for the existence of a cavity in the disk that is unresolved in our observations. The disk mass is set to 4.7~\msun{} and is modelled with $10^6$ particles. Two sink particles of 32~\msun{} and 3.2~\msun{} are introduced to represent the massive protostar in the disk and perturber, respectively. We refer to Ref.\citep{cuello2019} for the other parameters, and note that the minimum Toomre $Q$ parameter in the disk in our setup is 1.65, consistent with the observations.

The snapshot in Fig.~\ref{fig:flyby} is taken at about 12,600 years after the perturber passing by the periastron, using the visualization code \texttt{SPLASH}\citep{price2007}. The system has been rotated by 255\arcdeg{} and 52.8\arcdeg{} clockwise around the $z$ and $y$ axes, respectively.

We set up the simulation to capture the essential dynamical features of the flyby event, including the spirals, the tidal tail, and the path of condensation A, while we do not intend to reproduce all details of the observations. For example, we do not include disk warping, dust grain properties, or radiative transfer modeling, which may lead to specific differences such as the less extended tidal tail in the simulation. These differences do not affect the discussion of the essence of the flyby event. Detailed comparisons between such simulations and observations can be found in e.g.\ Ref.\citep{cuello2020}.

\vspace{0.2em}
\noindent\textbf{Likelihood of the flyby event in the observed environment}

We estimate the likelihood of the flyby event given the protostellar density in the environment of the Sagittarius C cloud. The problem of the flyby likelihood has been characterized in Refs.\citep{davies2011,pfalzner2013}. 

The key parameter is the (proto)stellar density in the considered area. Typical stellar densities at the central part of massive clusters are a few times $10^4$~pc$^{-3}$\citep{otter2021}. The stellar density in the massive star cluster forming region Sagittarius~B2(M) in the CMZ is estimated to be about 1$\times$$10^5$~pc$^{-3}$ \citep{ginsburg2018}. We thus adopt a value of 5$\times$$10^4$~pc$^{-3}$.

Then the time scale for a star to undergo a flyby event in this clustered environment can be estimated following\citep{davies2011,pfalzner2013}:
\begin{equation}
\tau\approx3.3\times10^7 \left(\frac{100~\textrm{pc}^{-3}}{n}\right)\left(\frac{v_\infty}{1~\kms{}}\right)\left(\frac{10^3~\textrm{AU}}{r_\textrm{peri}}\right)\left(\frac{\msun{}}{m}\right) \textrm{yr},
\end{equation}
where $n$ is the stellar density, $v_\infty$ is the mean relative velocity at infinity of the cluster stars, $r_\textrm{peri}$ is the periapsis of the flyby event, and $m$ is the total mass of the objects involved in the flyby event. The gravitational focusing effect, in which stars are deflected toward each other owing to the gravitational attraction between them, has been taken into account in the above equation.

We adopt $v_\infty=1$~\kms{}, $r_\textrm{peri}=2000$~AU, and $m=40$~\msun{} (the total mass of the disk, the central massive protostar, plus condensation A). The time scale is about 800 years. This flyby time scale is much shorter than the rotation period of the disk and the evolutionary time scale of the massive protostar in the disk ($10^4$ yr, equivalent to its Kelvin-Helmholz time scale). Therefore, it is highly likely that the disk undergoes one flyby event with a gas condensation of a few solar masses like condensation A during its evolution in this environment.

Finally, we note that this region is unique in that the disk system is the only massive object in the vicinity, and all the other objects that may be able to impact the dynamics of the disk are at most a few solar masses. The nearest external massive object is a 680~\msun{} gas core locating 0.25~pc away in projection to the southeast\citep{lu2019a}. Therefore, this highly mass-concentrated yet relatively isolated environment may have guaranteed the disk to grow sufficiently large and massive while be less affected by other objects except via close flybys.

\vspace{1em}
\noindent\textbf{Data availability}

This paper makes use of the following ALMA data: ADS/JAO.ALMA\#2018.1.00641.S. The data are available at \url{https://almascience.nao.ac.jp/aq} by setting the observation code. The reduced data used for this study are available from the corresponding authors upon reasonable requests.

\vspace{0.2em}
\noindent\textbf{Code availability}

The ALMA data were reduced using CASA versions 5.4.0 and 5.6.1 that are available at \url{https://casa.nrao.edu/casa_obtaining.shtml}. The code to make Fig.~\ref{fig:pd} is available at \url{https://doi.org/10.5281/zenodo.6413326}. The \texttt{3DBarolo} code is available at \url{https://github.com/editeodoro/Bbarolo}. The \texttt{PHANTOM} code is available at \url{https://github.com/danieljprice/phantom}. The \texttt{SPLASH} code is available at \url{https://github.com/danieljprice/splash}.

\vspace{0.2em}
\noindent\textbf{Acknowledgements}

We thank Hauyu Baobab Liu, Yu Cheng, and Patricio Sanhueza for helpful discussions. X.L.\ thanks his family, Qinyu E and Xiaoe Lyu, for their support during the preparation of this manuscript. X.L.\ acknowledges supports from the initial funding of scientific research for high-level talents at Shanghai Astronomical Observatory, and the Japan Society for the Promotion of Science (JSPS) KAKENHI grant No.\ 20K14528. G.-X.L.\ thanks Dr.\ Martin Krause for discussions on the flyby scenario. G.-X.L.\ acknowledges supports from NSFC grants W820301904 and 12033005. This work made use of the High Performance Computing Resource in the Core Facility for Advanced Research Computing at Shanghai Astronomical Observatory, and the Multi-wavelength Data Analysis System operated by the Astronomy Data Center (ADC), National Astronomical Observatory of Japan.
This work made use of the following ALMA data: ADS/JAO.ALMA\#2018.1.00641.S. ALMA is a partnership of ESO (representing its member states), NSF (USA) and NINS (Japan), together with NRC (Canada), MOST and ASIAA (Taiwan), and KASI (Republic of Korea), in cooperation with the Republic of Chile. The Joint ALMA Observatory is operated by ESO, AUI/NRAO and NAOJ.

\vspace{0.2em}
\noindent\textbf{Author contributions}

X.L.\ led the ALMA proposal, data reduction, numerical simulation, and paper writing. G.-X.L.\ led the interpretation of the data and analytical solutions, and contributed to numerical simulation and paper writing. Q.Z.\ commented on and helped to improve the article and the observing proposal. Y.L.\ contributed to the estimate of gas temperatures and commented on the article.

\vspace{0.2em}
\noindent\textbf{Competing interests}

The authors declare no competing interests.

\clearpage

\setcounter{figure}{0}
\renewcommand{\figurename}{Extended Data Fig.}
\renewcommand{\figureautorefname}{Extended Data Fig.}

\begin{figure}[!h]
\centering
\includegraphics[width=1\textwidth]{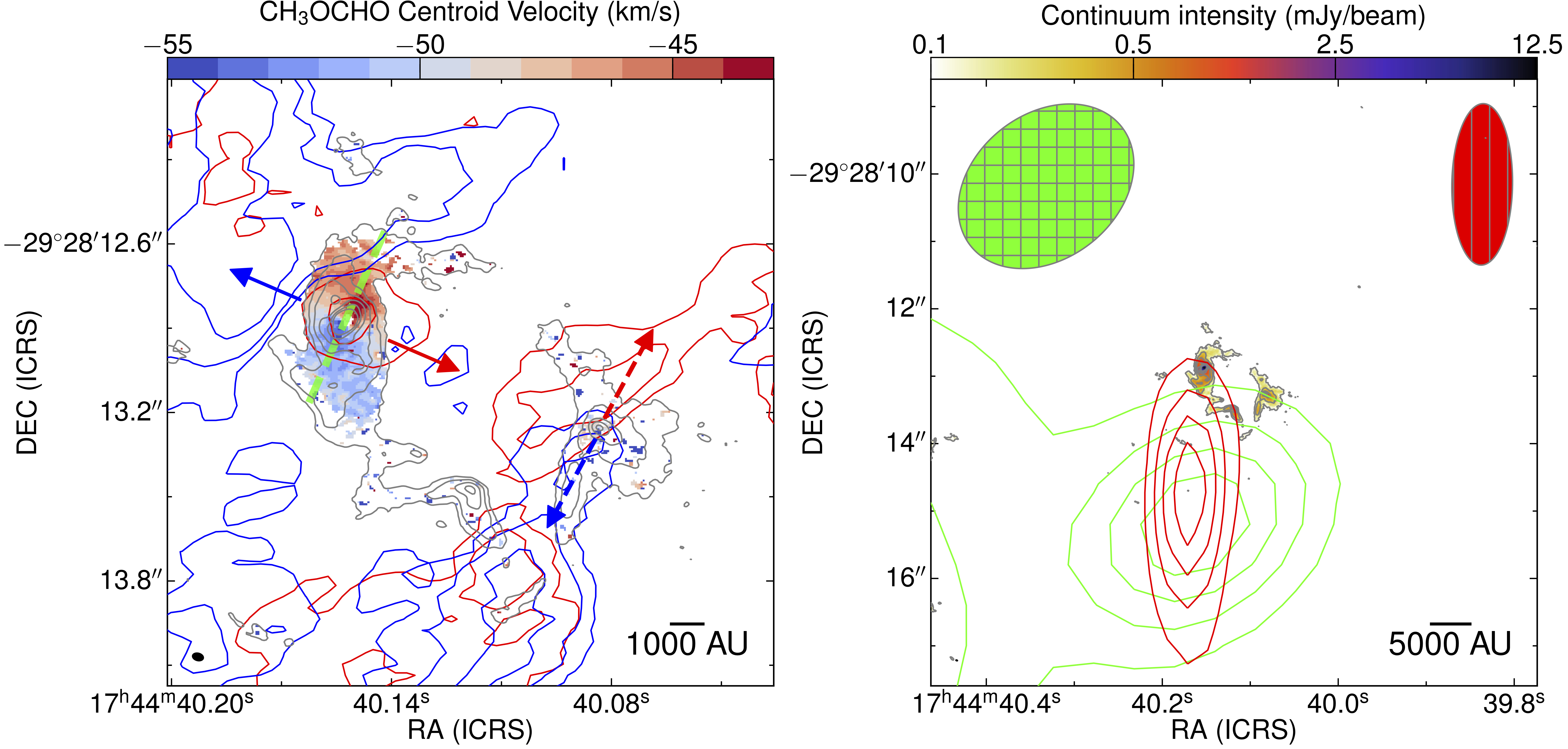}
\caption{
\textbf{Broader environment of the disk.} Left: the velocity field of the disk and the two condensations, derived from the \mf{} line. The blue and red contours show the blue and red-shifted SiO emission from previous ALMA observations\citep{lu2021}. The blue-shifted SiO emission is integrated between $-$80 and $-$51~\kms{}, and the red-shifted SiO emission between $-$48 and $-25$~\kms{}. The bipolar outflow associated with the disk, which has been identified in Ref.\citep{lu2021} using multiple molecular lines including SiO, is marked by the blue and red arrows. The best-fit kinematic major axis of the disk is denoted by the green dashed line, same as in Fig.~\ref{fig:disk}b. A candidate bipolar outflow associated with condensation A is marked by the dashed blue and red arrows. Right: the radio continuum emission in this region observed by the Very Large Array (VLA). Green contours are the 23~GHz continuum emission\citep{lu2019a}, while red contours are the 5.6~GHz continuum emission\citep{lu2019b}. The contour levels are between 20\% and 80\% and increment by 20\% of the peak intensity. The synthesized beams at the two frequencies are shown in the top left and top right corners, respectively. At both frequencies, the radio continuum emission is unresolved or marginally resolved. The background image and gray contours show the ALMA 1.3~mm continuum emission, same as in Fig.~\ref{fig:disk}.
}
\label{fig:largeenv}
\end{figure}

\begin{figure}[!h]
\centering
\includegraphics[width=1\textwidth]{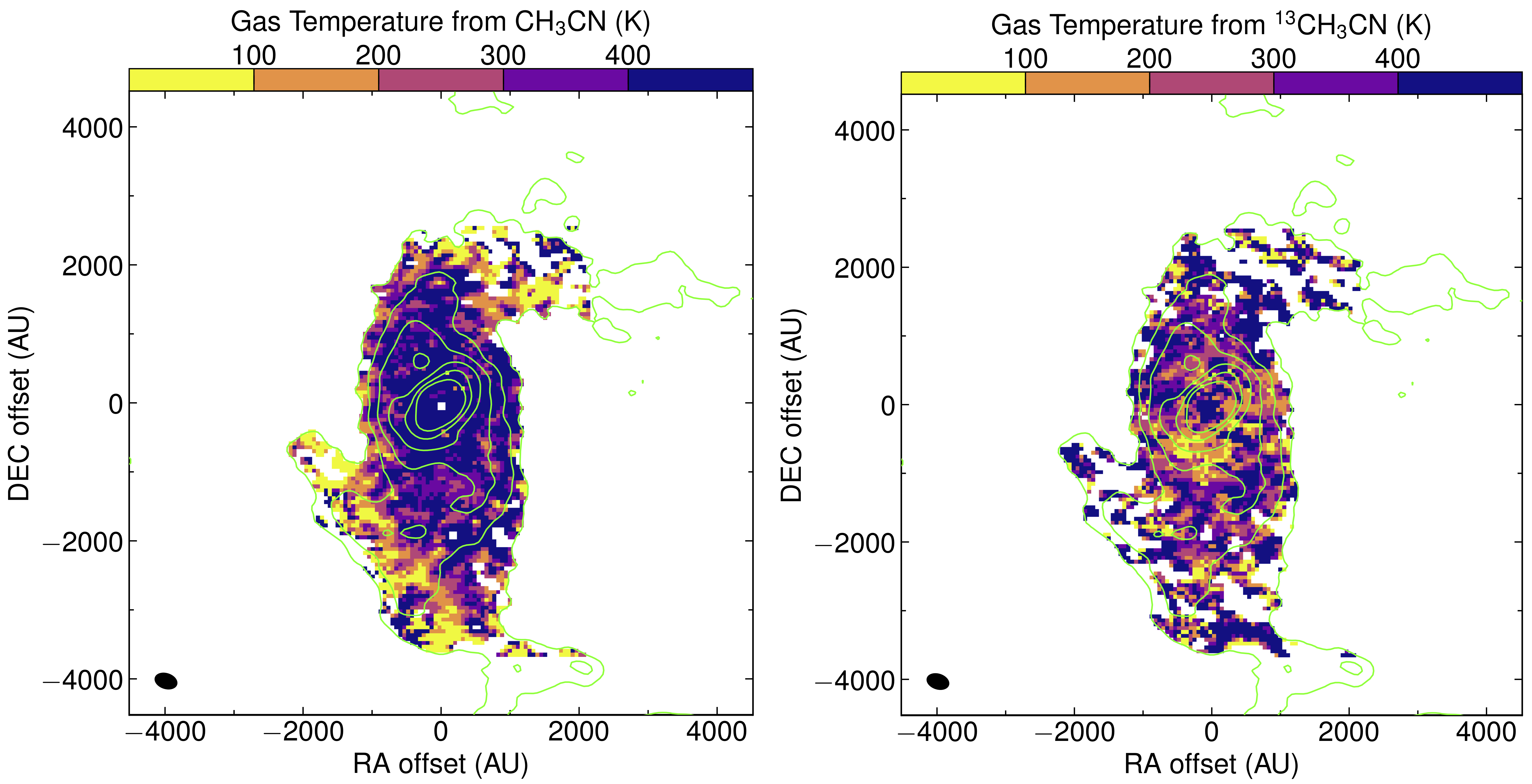}
\caption{
\textbf{Gas temperatures in the disk.} The two maps are derived from two groups of molecular lines: \mthc{} on the left, and \tmthc{} on the right, with the same scale range for comparison. The green contours show the continuum emission, with the same contour levels as in Fig.~\ref{fig:disk}a.
}
\label{fig:tgas}
\end{figure}

\begin{figure}[!h]
\centering
\includegraphics[width=1\textwidth]{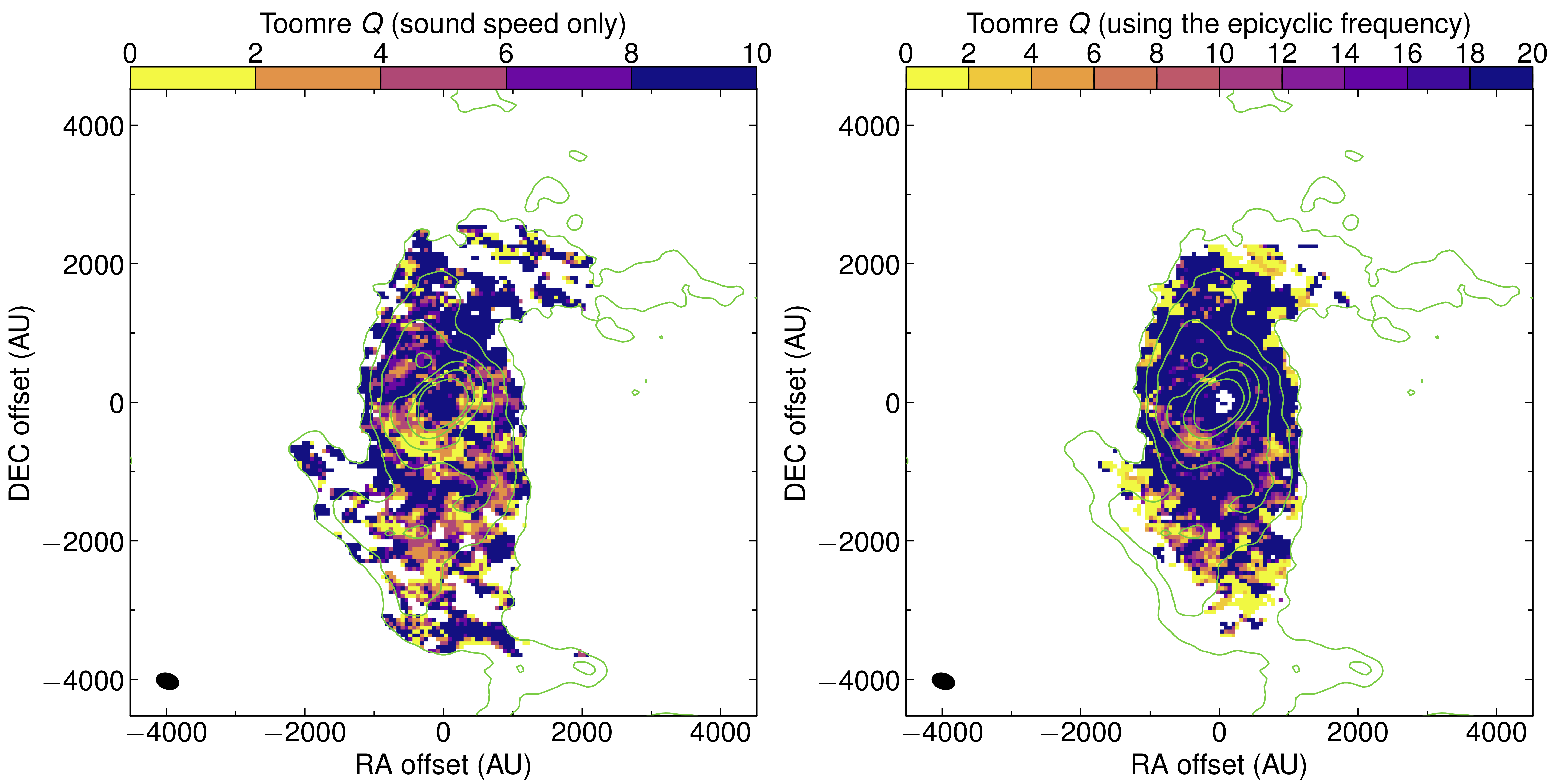}
\caption{
\textbf{Toomre $Q$ parameters in the disk with different assumptions.} Left: the case when using the sound speed only. Right: the case when using the epicyclic frequency following Equ.~\ref{equ:kappa}.
}
\label{fig:toomreQ_more}
\end{figure}

\begin{figure}[!h]
\centering
\includegraphics[width=1\textwidth]{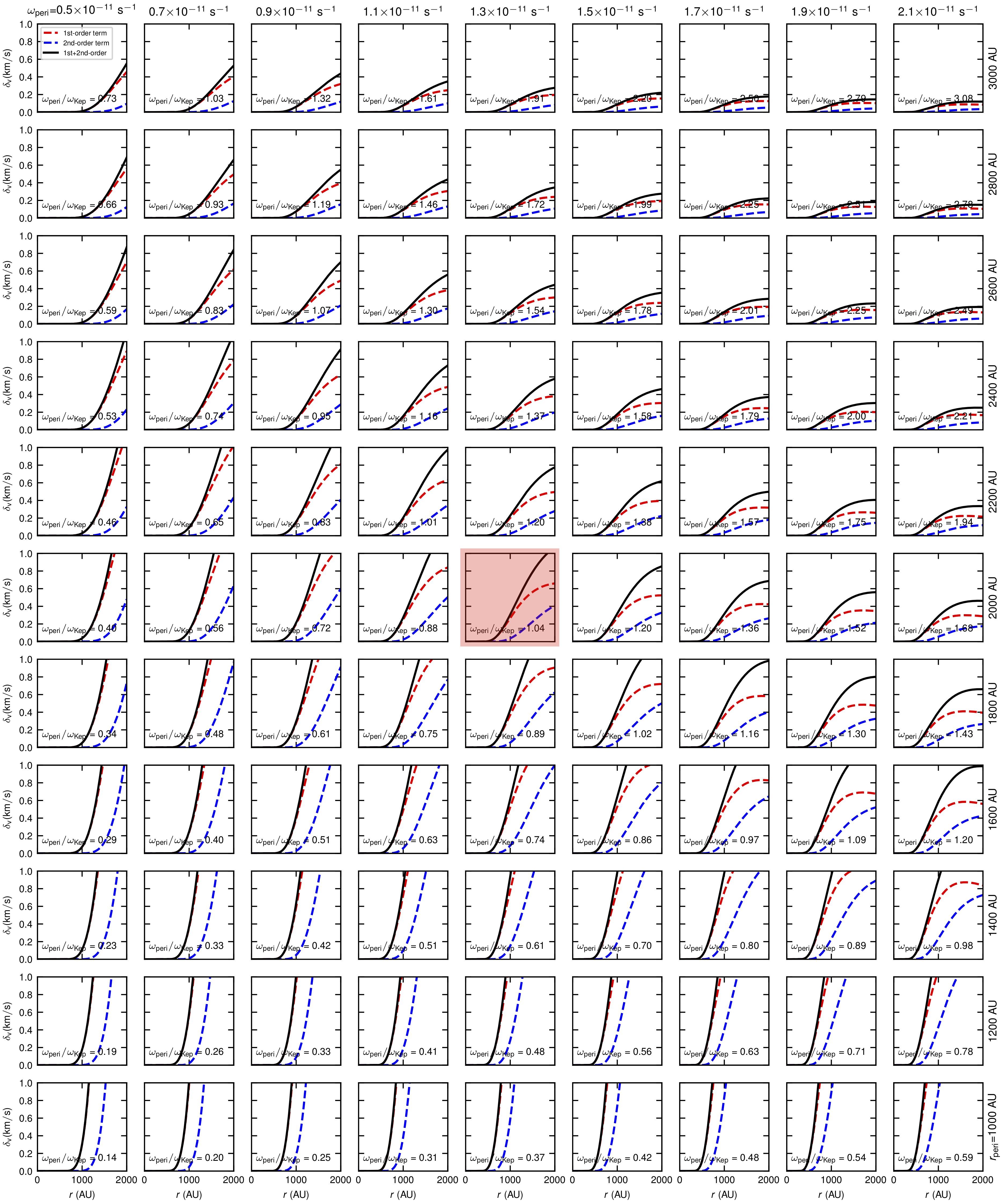}
\caption{
\textbf{A more extensive parameter space of perturbation-induced structure formation.} When the periastron distance is small (bottom panels), the perturber penetrates the disk and leaves a strong dynamical impact, thus truncating the disk. When large (top panels), the dynamical effect becomes insignificant. When the angular velocity at the periastron is low (left panels), the perturber is able to resonate only with outer radii of the disk that rotate more slowly. When high (right panels), the perturber resonates with the inner disk, disturbing smaller radii than observed. The only viable solution remains to the one identified in Fig.~\ref{fig:pd}.
}
\label{fig:pd_more}
\end{figure}

\clearpage

%\bibliography{sgrc.bib}
%\bibliographystyle{NA}

\end{document}